\documentstyle[12pt,twocolumn]{article}

\begin{document}

\setlength{\textwidth}{480pt}
\setlength{\textheight}{630pt}
\setlength{\topmargin}{-0.375in}
\setlength{\oddsidemargin}{-0.0833in}
\setlength{\evensidemargin}{-0.0833in}
\setlength{\parindent}{0in}
\setlength{\parskip}{6pt}
\setlength{\jot}{12pt}

\title{Electrodynamics and the Mass-Energy Equivalence Principle}

\author{
{\large Ezzat G. Bakhoum}\\
\\
{\normalsize University of West Florida}\\
{\normalsize P.O. Box 4200, Milton, FL. 32572 USA}\\
{\normalsize Email: bakhoum@modernphysics.org}\\
\\
{\normalsize (This work has been copyrighted with the Library of Congress)}\\
{\normalsize Copyright \copyright 2003 by Ezzat G. Bakhoum}
}

\date{\normalsize Originally posted on Oct. 3, 2003}

\maketitle

{\large\bf Abstract:}\\
\\
In this paper we investigate the link between classical electrodynamics and the mass-energy equivalence principle, in view of the conclusions reached in ref.\cite{Bakhoum1}. A formula for the radius of a charged particle is derived. The formula predicts the radius of the proton correctly. The radius of the electron turns out to be a surprising quantity that solves the existing problems of electrodynamics, particularly the problem of the infinite self-force of the electron. In addition, the classical radius of the electron (2.82fm) will prove to be not a ``radius", but rather the mean distance through which the retarded potentials of the self-force act. An important conclusion is that there is no deficiency in the classical Abraham-Lorentz model of the self-force, but rather the problem lies with our intuitive understanding of what an elementary particle is. Other important conclusions are also discussed, including a physically sound explanation for why electric charges must be quantized (as opposed to Dirac's monopole theory).\\
\\
{\large\bf 1. Introduction:}\\
\\
As the very successful theory of Quantum Electrodynamics (QED) was being developed during the three decades that started with Dirac's first paper on the subject in 1927 \cite{Schweber}, it was becoming clear that the major difficulty that the new theory must overcome is the inadequacy of the classical laws of electrostatics to describe electrodynamic phenomena at the quantum level. In 1948, for instance, J. Schwinger wrote\cite{Schwinger2} that the fundamental concepts of QED imply that ``only an accelerated electron can emit or absorb a light quantum"; hence, it was becoming clear during that early period that the classical Coulomb force of electrostatics has no existence if relative motion does not exist. That important fact was further supported and reinforced by the conclusions reached in ref.\cite{Bakhoum1}, in view of the total energy equation $H=mv^2$. Yet, despite the success of QED, no suitable explanation has emerged for why Coulomb's law (which is an undoubtedly accurate law) predicts the existence of an electrostatic force at rest. A seemingly unrelated and still unanswered question was about the physical sizes of the two most fundamental particles: the electron and the proton. Current experiments do unanimously agree that the radius of the proton is approximately 0.87fm \cite{Melnikov},\cite{Rosenfelder}, while similar experiments for measuring the radius of the electron have failed to agree on a unique number. The classical radius of the electron was known to be about 2.82fm for many years, until recent high energy collision experiments determined that the electron's radius cannot exceed $10^{-3}$fm; still, other experimental work by Dehmelt\cite{Dehmelt} and others showed that the radius of the electron must be on the order of $10^{-7}$fm. Currently, there is no experimental evidence to suggest other than that the electron is a structureless (point-like) Dirac particle. In connection with this issue was another, perhaps more important issue, which is the problem of the self energy of the electron that has plagued QED for many years and has necessitated the use of ``renormalization" techniques. Ironically, the renormalization techniques do not seem to have been taken seriously even by the founders of QED. Schwinger wrote in 1989 \cite{Schwinger1} that the problem of the self energy is ``possibly still unsolved", while R. Feynman wrote a few years earlier\cite{Feynman1} that ``I suspect that renormalization is not mathematically legitimate. What is certain is that we do not have a good mathematical way to describe the theory of quantum electrodynamics".\\
\\
It is our objective in this paper to demonstrate that all of the above problems are not only related, but in fact they are directly related to the topic that was treated in an earlier paper by the author\cite{Bakhoum1}, specifically, the misconception that had prevailed for nearly 100 years about what the mathematical relationship between mass and energy should be. This paper is organized as follows: in Sec.2 the total energy equation $H=mv^2$ will be used in the derivation of a classical mathematical formula for the radius of a charged particle. It will be demonstrated that the formula predicts the radius of the proton correctly. The radius of the electron will turn out to be a surprising quantity that solves the problem of the infinite self-force of the electron. The classical Abraham-Lorentz model of the self-force is also discussed in Sec.2. As we shall demonstrate, the classical radius of the electron, or 2.82fm, will prove to be not a ``radius" but rather the mean distance through which the retarded potentials of the self-force act. In Sec.3, the conclusions reached in Sec.2 will be applied to the case of the electron's wave function in the hydrogen atom; particularly, to the Dirac and the Schr\"{o}dinger models of the hydrogen atom. It will be demonstrated that the current knowledge about the electron's wave function in hydrogen further supports the conclusions reached in Sec.2. Finally, we will answer two fundamental -but very related- questions: first, why Coulomb's law of electrostatics predicts the existence of a force between two charged particles at rest, even though QED asserts that no such force can possibly exist at rest. Secondly, why electric charges always appear to be quantized, that is, appear as integer multiples of the fundamental electronic charge $q$.\\
\\
In our analysis, we will refer to the electron's charge as ``$q$" and to the quantity $q^2/4\pi\epsilon_0$ as ``$e^2$''. Extension of the conclusions reached here from the classical theory to the more elaborate theory of QED will be presented in a subsequent report. 

\pagebreak

{\large\bf 2. Dimensions of elementary particles and the divergence problem in electrodynamics:}\\
\\
{\bf 2.1. Radius of the electron:}\\
\\
The classical electromagnetic theory of the electron emerged as an extension of the radiation theories developed by Lorentz and Poincar\'{e} before the development of the theory of relativity. The connection between the radiation theories and the theory of the electron was the electromagnetic energy-momentum relationship. Before the discovery of relativity by Einstein, it was demonstrated that the momentum density $\vec{g}$ of the electromagnetic field is related to the time rate of flow of energy per unit area (or the Poynting vector), $\vec{S}$, by the following relation\cite{Feynman2}

\begin{equation}
\vec{g} = \frac{1}{c^2} \; \vec{S}.
\label{21}
\end{equation}

This relationship emerged as a direct consequence of the concept that the electromagnetic field carries the equivalent of a mass $m$ that is related to the energy $H$ stored in the field by the relationship $m=H/c^2$, which was again discovered classically before the introduction of the theory of relativity. By using $H=mc^2$ it is a very simple task to derive Eq.(\ref{21}). When Einstein introduced the principle of relativity in 1905 and again derived the equation $H=mc^2$, it was automatically assumed that Eq.(\ref{21}) must apply for material particles as well as electromagnetic energy. The classical theory of the electron was henceforth developed on the basis of that assumption.\\
\\
In the earlier paper by the author\cite{Bakhoum1} it was demonstrated that $H=mc^2$ is indeed valid {\em only} for the case of the electromagnetic field. For matter, the correct mass-energy equivalence relationship is $H=mv^2$. In this case it is again a very simple task to prove that

\begin{equation}
\vec{g} = \frac{1}{v^2} \; \vec{S}.
\label{22}
\end{equation}
  
We next consider the value of $\vec{S}$. According to classical electromagnetics, the Poynting vector $\vec{S}$ is given by

\begin{equation}
\vec{S} = \epsilon_0 c^2 \vec{E} \times \vec{B},
\label{23}
\end{equation}
 
where $\vec{E}$ and $\vec{B}$ are the traditional electric and magnetic field vectors at any point in space (note that the term $c^2$ here results from the application of Maxwell's equations). We note again that $\vec{S}$ represents the rate of flow of energy per unit area (note that the classical electromagnetic theory of the electron makes the assumption that the total energy of the electron is ``stored" in its electric and magnetic fields\cite{Feynman2}). If we now take the expression in Eq.(\ref{23}) and note, in addition, that there is a well-known relationship between the electric field $\vec{E}$ and the magnetic field $\vec{B}$ at any point in space in the vicinity of a moving particle, that second relationship being given by\cite{Feynman2},\cite{Eisberg} 

\begin{equation}
\vec{B} = \frac{1}{c^2} \; \vec{v} \times \vec{E},
\label{24}
\end{equation}

where $\vec{v}$ is the particle's velocity, we then have, from Eqs.(\ref{23}) and (\ref{24}),

\begin{equation}
\vec{S} = \epsilon_0 \vec{E} \times (\vec{v} \times \vec{E}).
\label{25}
\end{equation}

From Eqs.(\ref{22}) and (\ref{25}), the momentum density $\vec{g}$ will be given by

\begin{eqnarray}
\vec{g} & = & \frac{\epsilon_0}{v^2} \; \vec{E} \times (\vec{v} \times \vec{E})\nonumber\\
 & = & \frac{\epsilon_0}{|v|} \; \vec{E} \times (\hat{v} \times \vec{E})
\label{26}
\end{eqnarray}

where $\hat{v}$ is a unit vector in the direction of movement. The total momentum of the particle will be therefore given by

\begin{eqnarray}
\vec{p} & = & \frac{1}{|v|} \; \left[ \int d^3r \;\; \epsilon_0 \vec{E} \times (\hat{v} \times \vec{E}) \right]_{\mbox{\scriptsize radius}=a}^{\mbox{\scriptsize radius}=\infty}\nonumber\\
        & = & \frac{1}{|v|} \; \left[ \int d^3r \;\; (\epsilon_0 E^2 \sin^2 \theta) \;\; \hat{v}\right]_a^\infty 
\label{27}
\end{eqnarray}

where the symbol $a$ designates the particle's radius and where we have assumed that the sum of the momentum components in the direction normal to $\hat{v}$ will vanish. The integral in the above expression has already been evaluated in a number of references\cite{Feynman2},\cite{Jackson}. The result is $e^2/a$, with an optional coefficient that depends on the distribution of the charge within the volume defined by the radius $a$ (Jackson\cite{Jackson}, however, pointed out that if we carry the derivation in a Lorentz-covariant manner then the optional coefficient will be indeed 1). Accordingly, the expression in Eq.(\ref{27}) reduces to

\begin{equation}
\vec{p} = \frac{e^2}{|v| a} \; \hat{v}.
\label{28}
\end{equation}

As a scalar, the momentum $\vec{p}$ is equal to $mv$, i.e.,

\begin{equation}
|p| = mv = \frac{e^2}{v a}.
\label{29}
\end{equation}

Hence, the radius of the particle will be given by 

\begin{equation}
a = \frac{e^2}{mv^2}.
\label{31}
\end{equation} 

It is now very important to observe that this expression is similar to the one used to describe the classical radius of the electron (which is given by $e^2/m c^2$), but with $v^2$ instead of $c^2$ appearing in the denominator. Of course, the quantity $mv^2$ is just the total energy of the particle. It is also important to note that $m$ here must in general be assumed to be the relativistic mass of the particle, not the rest mass. We shall now proceed to examine the implications of the expression in Eq.(\ref{31}) in the specific case of a point-like Dirac  particle, such as the electron.\\
\\
{\bf 2.2. Self-energy of the electron:}\\
\\
It will be obvious what the implication of Eq.(\ref{31}) is when different velocities are considered. At $v=0$, the radius of the electron is infinite. At $v \approx c$, the relativistic mass $m$ approaches infinity, and hence the radius shrinks to zero. The physical meaning of this simple result is the following: the radius of an elementary particle is not a rigid quantity as previously thought. Since such a particle can be understood -within the framework of wave mechanics- as a wave function, then this result suggests that the spatial extension of the wave function is strongly correlated with the velocity. Indeed, this result is further confirmed by Heisenberg's uncertainty principle: if we can assert that the velocity of the particle is zero, then we must conclude that the spatial extension of the wave function is infinite. In fact, it is well known from the application of Schr\"{o}dinger's equation to describe the electron in certain configurations, such as in the hydrogen atom (more emphasis on that analysis is given later), that the electron's wave function can extend far beyond the distance known as the ``classical" radius of the electron. When, on the other hand, we consider the extreme of high velocities ($v \rightarrow c$), the wave function (and hence the radius) shrinks to zero as we indicated above. This result is of course nothing but the well-known concept of length contraction of special relativity. We must now point out that the contraction of the wave function at high velocities offers an alternative view of why it is impossible to accelerate any particle to a velocity beyond $c$. Since the wave function shrinks to zero, 
Coulomb interactions with the particle simply become impossible. The important conclusion here of course is that the theories of special relativity and wave mechanics are indeed very compatible theories, as long as the total energy of the particle is expressed correctly by the quantity $mv^2$.\\
\\
We shall now attempt to understand what the above result means for the problem of the self-energy of the electron, which is a well-known problem of electrodynamics that remains unsolved. As we know, the classical theory of electrodynamics assumes that the rest mass of the electron is due entirely to electromagnetic effects. According to the Abraham-Lorentz model of the electron, the rest mass is represented by the expression\cite{Feynman2}

\begin{equation}
m_0 = \frac{e^2}{a c^2},
\label{32}
\end{equation}

where $a$ is the electron's radius. Thus if $a \rightarrow 0$, then the ``rest energy" $m_0 c^2 \rightarrow \infty$. This is the well-known problem of ``divergence'' of the self-energy of the electron. Let us now examine what the implication of the corrected expression of the radius $a = e^2 / mv^2$ (Eq.\ref{31}) is when the electromagnetic rest mass is concerned. Obviously, the independent variable in this case is the velocity $v$. As we indicated above, if $v \approx 0$ then $a \rightarrow \infty$ and the total energy $m_0 v^2 \rightarrow 0$. Accordingly, the divergence problem is eliminated at small velocities. Of course, the rest mass $m_0$ remains finite in this case. As we also indicated above, if $v \rightarrow c$, then $m \rightarrow \infty$ and the radius $a \rightarrow 0$, which are just of course the relativistic predictions.
The important result here is that the divergence problem is eliminated at small velocities.\\
\\
Another important method for calculating the self-energy of the electron is to calculate the energy stored in the electric and magnetic fields generated by the particle. This energy must still be given by the total energy expression $H=mv^2$, and, in terms of fields, will be given by\cite{Jackson}

\begin{equation}
\frac{1}{2} \; \left[ \int d^3r \;\; (\epsilon_0 E^2 + \frac{1}{\mu_0} B^2)  \right]_{\mbox{\scriptsize radius}=a}^{\mbox{\scriptsize radius}=\infty}
= m v^2.
\label{34}
\end{equation}  

Normally, that integral is computed between the limits $a$ and infinity. It is well known mathematically that if the radius $a$ is taken to be equal to zero the integral diverges. Given the conclusions that we just reached, however, we now understand that $a$ will be equal to zero only at very high velocities. At low velocities, $a \rightarrow \infty$ as we concluded and hence the numerical value of the integral will be approximately zero. We can now see that this formulation for the total energy (that is, in terms of fields) is numerically in agreement with the quantity $mv^2$ on the r.h.s. of Eq.(\ref{34}).\\
\\
{\bf 2.3. A better understanding of the Abraham-Lorentz force:}\\
\\
Our objective now is to understand the classical Abraham-Lorentz model of the electron in view of the conclusions reached above. In order to better understand the Abraham-Lorentz model, we must first understand in general how the derivation was performed. The self-force is represented by the classical Lorentz force equation\cite{Feynman2},\cite{Jackson}

\begin{equation}
F_{\mbox{self}} = - \int d^3 r \;\; (\rho \vec{E_s} + 
\vec{J} \times \vec{B_s})
\label{35}
\end{equation}

where $\vec{E_s}$ and $\vec{B_s}$ are the self-fields of the electron and where the negative sign indicates that the force will be in the opposite direction to the external force acting on the particle. It is not difficult to verify, in view of Eq.(\ref{24}), that the second term in the integrand contains terms on the order of $v^2/c^2$. If the electron is non-relativistic (that is, $v \ll c$), then it is reasonable to neglect the second term. The self force will then be given by

\begin{equation}
F_{\mbox{self}} = \int d^3 r \;\; \rho (\nabla \phi + 
\frac{\partial \vec{A}}{\partial t}),
\label{36}
\end{equation}

where

\begin{eqnarray} 
\phi (x,y,z,t) & = & \int \frac{ \rho(x^\prime,y^\prime,z^\prime,t^\prime)_{ret}}{4\pi\epsilon_0 R} \;\; 
dx^\prime dy^\prime dz^\prime \nonumber\\
\vec{A} (x,y,z,t) & = & \int \frac{ \vec{J}(x^\prime,y^\prime,z^\prime,t^\prime)_{ret}}{4\pi\epsilon_0 c^2 R} \;\; 
dx^\prime dy^\prime dz^\prime \nonumber\\
 & & 
\label{37}
\end{eqnarray}

are the scalar and vector potentials. The details of the integration are quite elaborate, but the result is a Taylor series of the form

\begin{equation}
F_{\mbox{self}} = K_1 \frac{e^2}{R c^2} \ddot{x} - \frac{2}{3} \frac{e^2}{c^3} \frac{d^3 x}{dt^3} + K_2 \frac{e^2 R}{c^4} \frac{d^4 x}{dt^4} + \ldots
\label{38}
\end{equation}

where $K_1$ and $K_2$ are numerical constants. In this classical expression, $K_1 = 2/3$ due to the fact that the theory in not Lorentz covariant in form. However, it has also been demonstrated\cite{Jackson} that a Lorentz covariant form of the theory will yield $K_1 = 1$. Eq.(\ref{38}) is the classical Abraham-Lorentz expression of the self-force. As it is well known, the whole difficulty behind the divergence problem in this classical theory as well as in QED lies with the numerical value that we must assign to the distance $R$. In the above derivation, $R$ must be strictly understood as the average distance through which the retarded potentials $\phi$ and $\vec{A}$ act. Intuitively, if we think of the electron as a rigid, charged sphere, then it is very reasonable to assume that $R=a$, or the radius of the particle. But if we reach the conclusion that the radius $a$ can be very large (in fact, infinite if $v \approx 0$), then how $R$ can be understood? In the expressions in Eq.(\ref{37}), the distance $R$ is the average distance through which the retarded potentials propagate from a charge $\rho(x^\prime,y^\prime,z^\prime,t^\prime)$ to another charge $\rho(x,y,z,t)$ such that the two charges are inside each-other's light cone (we have to stress here that the classical model is not relativistic, and hence lacks Lorentz covariance, but this statement is generally true). $R$ then must be a distance that guarantees that the necessary condition of Lorentz invariance is not violated. Another obvious problem with a very large radius is the shielding effect of a group of charges upon any given single charge. In other words, any given charge $\rho(x^\prime,y^\prime,z^\prime,t^\prime)$ can only communicate its retarded potentials to a limited number of other charges that effectively form a ``shield" around that charge. The mean distance through that shield must obviously be finite and limited. Given these two important restrictions, we have no choice but to conclude that the average distance $R$ {\em is not} the radius of the particle. The question now is, what is the numerical value of $R$? Since the classical theory suggests (and in fact the theory of QED asserts) that all the observed mechanical mass $m_0$ of the electron is electromagnetic in origin, then the term $e^2 / R c^2$ in Eq.(\ref{38}) must represent that mass. Therefore we have

\begin{equation}
R = \frac{e^2}{m_0 c^2} = 2.82 \; \mbox{fm}
\label{39}
\end{equation}

This famous result was well known before the theory of relativity was discovered and was assumed to represent the ``radius" of the electron (note that here, again, the term $c^2$ had emerged from the application of Maxwell's equations, not from special relativity). It is now clear that this numerical value of 2.82fm does not represent a ``radius", but it is actually the mean distance through which the retarded potentials propagate, given the physical restrictions mentioned above. The actual radius of the electron will be as given by Eq.(\ref{31}).\\
\\
{\bf 2.4. Radius of the proton:}\\
\\     
As we know, the proton is not a point-like Dirac particle but is composed of partons (the quarks). The wave function of such a composite particle, therefore, can be described as an eigenstate of two fields: a near-field and a far-field. From a large distance, the proton will be seen as a point-like particle and its radius will be perceived as described  above. This is the far-field radius. At a very short distance, however, the radius will be modified by an uncertainty term that is caused by the movement of the partons. This is the near-field radius and is the one that is usually measured in high-energy collision experiments. At high energies, the radius given by Eq.(\ref{31}) will be approximately zero as we indicated earlier. The near-field radius of the proton will be therefore $a + \Delta a$, where $a \approx 0$. The task now is to determine the value of the uncertainty term $\Delta a$ that is caused by the movement of the quarks. We shall verify that this value matches precisely the modern value of 0.87fm, which is known as the ``radius of the proton"\cite{Melnikov},\cite{Rosenfelder}.\\
\\
Using Eq.(\ref{31}), the uncertainty in the radius $\Delta a$ can be written as follows

\begin{equation}
\Delta a = \frac{e^2}{\Delta (mv^2)} = 
\frac{e^2}{\Delta p \Delta v},
\label{40}
\end{equation}

where $\Delta (mv^2)$ is the uncertainty in the total proton's energy and where $\Delta p$ and $\Delta v$ are the uncertainties in the momentum and velocity, respectively. The jet experiments at DESY\cite{Desy} and CERN\cite{Cern} in the early 1980s showed that the transverse hadrons ($q\bar{q}$ pairs) resulting in high-energy proton-antiproton collision experiments had a transverse momentum of 450Mev/$c$ on the average; which means an average transverse momentum of 225Mev/$c$ for a single quark. We will now assume that this value, 225Mev/$c$, is the value of $\Delta p$ that we are seeking. The remaining task is to find the value of $\Delta v$. Following the guidelines of QED, the uncertainty in the quark's velocity $\Delta v$ must be on the order of $\alpha c$, where $\alpha$ is the fine structure constant; that is,   

\begin{equation}
\Delta v = \alpha c = \left( \frac{e^2}{\hbar c} \right) c = \frac{e^2}{\hbar} = \frac{c}{137}.
\label{41b}
\end{equation}

Substituting for the values of $\Delta p$ and $\Delta v$ in Eq.(\ref{40}) gives 

\begin{eqnarray}
\Delta a & = & \frac{e^2}{(225 \mbox{Mev}/c) \cdot (c/137)} = 
\frac{e^2}{\frac{225}{137} \mbox{Mev}} \nonumber\\
 & = & \frac{q^2 / 4 \pi\epsilon_0}{1.64 \mbox{Mev}} \nonumber\\
 & = & 0.875 \mbox{fm}
\label{41c}
\end{eqnarray}

This number is of course the currently accepted value of the proton's radius. It must be therefore now clear that this value represents not the proton's radius, but more accurately the uncertainty in the radius due the motion of the partons. It is also interesting to note, from Eq.(\ref{40}), that

\begin{equation}
\Delta a \Delta p = \frac{e^2}{\Delta v} = \frac{e^2}{e^2/ \hbar} = \hbar
\label{41}
\end{equation}

which is just of course Heisenberg's uncertainty principle. Accordingly, it must be further clear that Eq.(\ref{40}) is in agreement with the known laws of quantum mechanics.\\
\\
\\
{\large\bf 3. The electron in the hydrogen atom and a deeper understanding of the Coulomb force:}\\
\\
There is a very interesting observation to be made about the radius of the electron in a Bohr orbit. In a Bohr orbit, the electron is in equilibrium due to the equality of the Coulomb and the centrifugal forces acting on it, that is,

\begin{equation}
\frac{e^2}{R^2} = \frac{mv^2}{R},
\label{42}
\end{equation}  

where $R$ is the radius of the orbit. Therefore $R = e^2/mv^2$. But this quantity is numerically the same as the radius $a$ of the electron given by Eq.(\ref{31}). Hence, an electron in a Bohr orbit has a radius that is precisely equal to the radius of the orbit! The question now is why this is the case? Before we answer that question, however, we will first demonstrate that this simple result is indeed distinguishable not just from Bohr's model, but further from Dirac's model of the hydrogen atom. A full discussion of the Shr\"{o}dinger model will be given in a subsequent paper. In Sec. 3.2, the relationship between this result and Coulomb's law of force will be discussed and a very important physical limitation on Coulomb's law that was previously unknown will be shown. Finally, we will reach a physically sound explanation for why electric charges always appear to be quantized. \\
\\
{\bf 3.1. Relationship between the electron's radius and Dirac's model of the hydrogen atom:}\\
\\
For the electron in the hydrogen atom, Dirac\cite{Dirac} had proposed that the electron's wave function $\psi$ must be an eigenstate of two radial functions $\psi_a$ and $\psi_b$, such that 

\begin{equation}
\psi_a = f \: e^{- \gamma R}, \qquad \psi_b = g \: e^{- \gamma R},
\label{43}
\end{equation}

where $\gamma$ is a constant and where $f$ and $g$ are power series of the form  $f= \sum_s C_s R^s$, $g= \sum_s C_s^\prime R^s$. Moreover, the general assumption is that one of the two wave functions dominates greatly over the other, hence,

\begin{equation}
\psi \approx (\sum_s C_s R^s) \: e^{- \gamma R}
\label{44}
\end{equation}

The question now is how can we relate the physical dimensions of the electron to this representation of its wave function? Dirac pointed out that the series

\begin{equation}
f = C_{s_0} R^{s_0} + C_{s_1} R^{s_1} + \ldots + C_{s_n} R^{s_n}
\label{45}
\end{equation}

must converge for small $R$ as well as for large $R$. At the point where $R=a$ (the radius of the particle), the series must therefore vanish. We will take as a special case the case of the hydrogen atom in the ground state to see if in fact the radius of the electron is equal to the radius of the orbit. For the ground state, $n=1$ and $s_n=n+s_0=1+s_0$ (see also the analysis in ref.\cite{Bakhoum2}). Hence, at the point where $R=a$, we must have the condition

\begin{equation}
f(a) = C_{s_0} a^{s_0} + C_{s_1} a^{1+s_0} = 0.
\label{46}
\end{equation}

Solving for $a$ gives $a= - C_{s_0}/C_{s_1}$. Dirac\cite{Dirac} did in fact demonstrate that the ratio $- C_{s_0}/C_{s_1}$ is equal to $\hbar^2 / me^2 =   e^2/mv^2$ (note that $v=e^2/\hbar$ in a Bohr's orbit). In other words, Dirac proved that the ratio of the first two constant coefficients in the series is numerically equal to Bohr's orbit (0.53 Angstroms). But we just demonstrated that the electron's wave function will vanish precisely at the point where $R = - C_{s_0}/C_{s_1}$; hence, the electron's wave function must be contained within a radius of 0.53 Angstroms. This confirms the result reached previously in the case of the simple Bohr model.\\
\\
{\bf 3.2. The concept of ``Force" in electrodynamics versus electrostatics:}\\
\\
As we indicated earlier, since the early days of the theory of QED it became known that no exchange of force between two charged particles is possible unless relative motion between the particles is present; yet, to date, no suitable explanation has emerged for why Coulomb's law predicts the existence of an electrostatic force at rest. The important question that must then be addressed is: how the concept of ``force" should be treated mathematically within the modern theories of electrodynamics? In the classical theory of electrostatics, the electric field intensity $\vec{E}$ is defined as the force exerted on a unit charge placed in that field. The terms ``electric field intensity" and ``force" are therefore two equivalent statements. Feynman was one of the few QED theorists who realized early on that the concepts of ``force" and ``field" must be separated\cite{Schweber}. As we know today, in macroscopic electrostatics, the force that exists between two charged objects is in reality due to the exchange of quantum force carriers, or virtual photons, between electrons which are actually in a state of relative motion (that is, in atomic orbits). This fact wasn't of course known to Coulomb when he discovered his famous inverse square law of electrostatics. Only in the 1940s it became accepted that the existence of a ``field" between two charged particles does not mean the existence of a ``force" and it was then when Schwinger wrote\cite{Schwinger2} that ``only an accelerated electron can emit or absorb a light quantum". Hence, it became clear that the electrostatic ``field" is an abstract concept that provides a mathematical tool for calculating the Coulomb force, but does not by itself constitute a physical entity that exists at every point in space as classical electrostatics suggests. Even though Feynman and Schwinger realized this fact, it was still very difficult to eliminate the field concept entirely, due to its necessity in making the calculations.\\
\\
It is our objective in this section to demonstrate that not just the electrostatic field is an abstract concept, but so is the concept of the electric ``charge". Indeed, we will demonstrate that if we combine classical mechanics with the electromagnetic mass of the electron according to Eq.(\ref{31}), Coulomb's force equation is generated directly, thus eliminating both of the concepts of ``field" and ``charge". A much better understanding of the laws of electrostatics and particularly of the Coulomb force will then emerge.\\
\\
We will consider first the motion of an electron of a charge $q$ in the vicinity of a larger positive charge $Q$. For simplicity we will consider only the case  of a stable orbit (e.g., circular). By using Eq.(\ref{31}) to represent the electromagnetic mass as a function of the radius, the centrifugal force acting on the electron will be given by

\begin{equation}
F = m \frac{v^2}{R} = \left( \frac{e^2}{a v^2} \right) \; \frac{v^2}{R} = 
\frac{e^2}{a R} = \frac{q^2}{4\pi\epsilon_0 \; a \; R},
\label{58}
\end{equation} 

The expression in Eq.(\ref{58}) is an electrostatic force equation that is similar, but not mathematically identical, to the Coulomb force. Since we have no experimental evidence to suggest that Coulomb's law is invalid, we ask, how Coulomb's law can be obtained from the above expression? Obviously, if we choose $a = R/Z$, where $Z$ is an integer, then we have

\begin{equation}
F = \frac{Z \; q^2}{4\pi\epsilon_0 \; R^2},
\label{59}
\end{equation} 

which is the commonly known form of Coulomb's law. There is, however, an important difference between this formula and Coulomb's law. This difference is not a mathematical difference but rather a conceptual one: at rest, the radius of the electron $a \rightarrow \infty$, hence we must conclude that $Z \rightarrow 0$ and therefore $F = 0$ at rest. Accordingly, the important difference between Eq.(\ref{59}) and Coulomb's law is that $Z$ must be allowed to have the numerical value of $0$!\\
\\
It is already clear from this expression, since Coulomb's law has been generated automatically without the necessity for a field equation, that the concept of ``field" is indeed an abstract concept. Let us now examine how we can also make the conclusion that the electric ``charge" is an abstract concept. As the QED theorists knew, the carrier of the Coulomb force is the quantum of electromagnetism, that is, the photon. The conclusion that emerged during the 1940s that no light quanta are exchanged at rest now becomes, in view of the above, equivalent to: ``no light quanta are exchanged when the integer $Z=0$". In this case, we ask the question: what does the integer $Z$ represent? It must represent the number of light quanta exchanged. For instance, two charges of magnitudes $Z_1 q$ and $Z_2 q$ will exchange $Z = Z_1 Z_2$ light quanta. At rest, no light quanta are exchanged and hence $Z=0$. This conclusion therefore gives Coulomb's law its usual mathematical expression (Eq.\ref{59}) in addition to explaining why no force exists at rest.\\
\\
We can now see that the abstract concept of the ``field", defined mathematically by Gauss' law as $E = q/4\pi\epsilon_0 R^2$, had to be combined with the abstract concept of the ``charge", defined mathematically as $Q=Zq$, in order to generate Eq.(\ref{59}). These are the concepts that we are used to in classical electrostatics. The underlying physics, however, can be simply stated as follows: the electrostatic force $F$ between any two ``charges" is always quantized in units of $e^2/R^2$. The quantum number $Z$ is the number of light quanta, or force carriers, exchanged between the two charges (note that the energy content of each light quantum has to be equal to $e^2/R$). This explains why any conventional charge always appears ``quantized". Clearly, the quantization of the force, because of the discrete nature of the photons, is the more fundamental concept. Hence, it is now clear that the concepts of ``field" and ``charge" are both abstract concepts indeed.\\
\\
Perhaps the only concept that is radically new here is the concept of the quantization of the radius. But we recall again that the radius essentially describes the spatial extent of the wave function. Accordingly, the quantization of both the force (in units of $e^2/R^2$) and the spatial extension of the wave function is all what is needed to describe a Coulomb interaction. It is important to note that in the case of a non-stable orbit the radius {\em will not} be quantized as described above; that is, the force in Eq.(\ref{58}) can actually be larger or smaller than the Coulomb force.\\
\\
When we apply the force expression given by Eq.(\ref{59}) to the case of the hydrogen atom (stable orbit), we note that $Z=1$ for hydrogen. Since we have assumed that $a = R/Z$, then we must conclude that $a=R$; that is, the radius of the electron in the hydrogen atom is precisely equal to the radius of the orbit. This is of course the conclusion that was reached previously from the Bohr model.\\
\\
\\
{\large\bf 4. Summary:}\\
\\
By using the total energy equation $H = mv^2$ the outstanding problems of electrodynamics are essentially eliminated. We have demonstrated that the radius of an elementary particle such as the electron is not a ``rigid" quantity as previously thought, but is rather a function of the velocity. This conclusion results in the successful elimination of the ``divergence" problem of the self force of the electron. In addition, it indicates that the concept of ``radius" is essentially a description of the spatial extent of the wave function. Moreover, the well-known distance of 2.82fm, long believed to represent the radius of the electron, proved to be not a radius but rather the mean distance through which the retarded potentials of the self force act.\\
\\
Important conclusions about the physical dimensions of the electron in the hydrogen atom were made. Essentially, we have concluded that the electron's wave function must be totally confined within the radius defined by the Bohr orbit. This conclusion was verified from the Bohr and Dirac models of the hydrogen atom. Moreover, if we introduce the postulate that the radius of the electron in a Coulomb interaction must be quantized, this directly leads to the Coulomb force equation without reliance on the classical laws of electrostatics. In view of this, two other important conclusions were made: first, that the integer  $Z$ in the Coulomb force equation must have the numerical value of 0 at rest. Secondly, the concept of the ``electric charge" proved to be an abstract concept, just like the ``field" concept. In reality, what occurs in a Coulomb interaction is an exchange of a number $Z$ of virtual photons, or force carriers, where the energy content of each force carrier is  always equal to $e^2/R$, regardless of the physical nature of the particle emitting that carrier. It is then due to the quantum nature of this fundamental boson of electromagnetism that any charge must appear to be ``quantized". Hence, the quantization of the force, not the charge, is the more fundamental physical description of the Coulomb interaction.


\end{document}